\documentclass[a4paper,11pt]{article}

\pdfoutput=1 

\usepackage{jheppub} 
\usepackage[bottom]{footmisc}
\usepackage{amssymb}
\usepackage{amsmath}
\usepackage{amsthm}
\usepackage[usenames,dvipsnames]{xcolor}
\usepackage{epsfig}
\usepackage{dcolumn}
\usepackage{tikz}
\usetikzlibrary{shapes.geometric, arrows}
\usepackage{upgreek}
\usepackage{setspace}
\usepackage{enumitem}
\usepackage{array,multirow,bigdelim,arydshln}
\usepackage{appendix}
\usepackage{xparse}
\usepackage{stmaryrd}
\usepackage[T1]{fontenc} 
\usepackage{mathtools}
\usepackage{physics} 
\usepackage{adjustbox}
\usepackage{multirow}
\usepackage{graphicx} 
\usepackage{float} 
\graphicspath{{./images/}}
\usepackage[nottoc]{tocbibind}
\usepackage{hyperref}
\usepackage[utf8]{inputenc}
\hypersetup{
	colorlinks,
	urlcolor=Maroon,
	linkcolor=Maroon,
	citecolor=Maroon
	}

\NewDocumentCommand{\binomial}{omm}
 {%
  \genfrac(){0pt}{}{#2}{#3}%
  \IfValueT{#1}{_{\!#1}}%
 }
\NewDocumentCommand{\eulerian}{omm}
 {%
  \genfrac<>{0pt}{}{#2}{#3}%
  \IfValueT{#1}{_{\!#1}}%
 }

\def \s {\sigma}

\usepackage{latexsym}
\usepackage{tikz}

\theoremstyle{plain}

\newtheorem{theorem}{Theorem}[section]
\newtheorem{thm}{Theorem}[section]

\newtheorem{cor}[thm]{Corollary}
\newtheorem{defn}[thm]{Definition}
\theoremstyle{definition}

\usepackage[percent]{overpic}
\usepackage{multirow} 
\usepackage{slashed}

\title{Planar 
Matrices and 
Arrays of Feynman Diagrams: Poles for Higher $k$ }

\author[a,b,c]{Alfredo Guevara}
\emailAdd{aguevaragonzalez@fas.harvard.edu}
\author[d,a,e]{and Yong Zhang}\emailAdd{yzhang@perimeterinstitute.ca}

\affiliation[a]{Perimeter Institute for Theoretical Physics, Waterloo, ON N2L 2Y5, Canada}
\affiliation[b]{Department of Physics \& Astronomy, University of Waterloo, Waterloo, ON N2L 3G1, Canada}
\affiliation[c]{Society of Fellows, Harvard University, Cambridge, MA 02138, USA}
\affiliation[d]{CAS Key Laboratory of Theoretical Physics, Institute of Theoretical Physics, Chinese Academy of Sciences, Beijing 100190, China}
\affiliation[e]{School of Physical Sciences, University of Chinese Academy of Sciences, No.19A Yuquan Road, Beijing 100049, China}

\abstract{Planar arrays of tree diagrams were introduced as a  generalization of Feynman diagrams that enables the computation of biadjoint amplitudes $m^{(k)}_n$ for $k>2$ . In this follow-up work we investigate the poles of $m^{(k)}_n$ from the perspective
of such arrays. For general $k$ we characterize the underlying polytope as a Flag Complex and propose a computation of the amplitude based solely on the knowledge of poles, whose number is drastically less than the number of full arrays. As an example we first provide all the poles for the cases $(k,n)=(3,7),(3,8),(3,9),(3,10),(4,8)$ and $(4,9)$ in terms of 
their planar arrays of degenerate Feynman diagrams.
We then implement a simple compatibility criteria together with an addition operation between arrays, and recover the full collections/arrays 
for such cases. Along the way we implement hard and soft kinematical limits, which provide a map between poles in kinematic space and their combinatoric arrays. We use the operation to give a proof of a previously conjectured combinatorial duality for arrays in $(k,n)$ and $(n-k,n)$. We also outline the relation to boundary maps of the hypersimplex $\Delta_{k,n}$ and rays in the tropical Grassmannian $\textrm{Tr}(k,n)$. 

}

\begin{document}
\maketitle
\addtocontents{toc}{\protect\setcounter{tocdepth}{1}}
\def \tr {\nonumber\\}
\def \la  {\langle}
\def \ra {\rangle}
\def\hset{\texttt{h}}
\def\gset{\texttt{g}}
\def\sset{\texttt{s}}
\def \be {\begin{equation}}
\def \ee {\end{equation}}
\def \ba {\begin{eqnarray}}
\def \ea {\end{eqnarray}}
\def \k {\kappa}
\def \h {\hbar}
\def \r {\rho}
\def \l {\lambda}
\def \be {\begin{equation}}
\def \en {\end{equation}}
\def \bes {\begin{eqnarray}}
\def \ens {\end{eqnarray}}
\def \red {\color{Maroon}}
\def \pt {{\rm PT}}
\def \s {\sigma} 
\def \ls {{\rm LS}}
\def \ma {\Upsilon}
\def \s {\textsf{s}}
\def \t {\textsf{t}}
\def \R {\textsf{R}}
\def \W {\textsf{W}}
\def \U {\textsf{U}}
\def \e {\textsf{e}}
\def \gr {\text{Gr}}
\def \trg {\text{TrG}}

\numberwithin{equation}{section}

\section{Introduction}

The Cachazo-He-Yuan (CHY) formulation provides a direct window into the scattering amplitudes of a wide range of Quantum Field Theories, by expressing them as a localized integral over the moduli space of punctures of $\mathbb{CP}^1$ \cite{Fairlie:1972zz,Fairlie:2008dg,Cachazo:2013gna,Cachazo:2013hca,Dolan:2013isa}. Such formulation was generalized by Cachazo, Early, Mizera and one the authors (CEGM), who extended it to configuration spaces over $\mathbb{CP}^{k-1}$, or equivalently to the Grasmanian $\gr(k,n)$ modulo rescalings \cite{Cachazo:2019ngv}. This unveiled a beautiful connection to tropical geometry, revealing that the CEGM amplitudes (for the generalized biadjoint scalar theory, $m^{(k)}_n$) can be computed either from a CHY formula or by more geometrical methods \cite{Cachazo:2019ngv,Drummond:2019qjk,Cachazo:2020uup}. In particular, the full amplitude $ m^{(k)}_n  ( \mathbb{I}|\mathbb{I})$ can be obtained as the volume of the positive Tropical Grassmannian $\trg^ +(k,n)$ viewed as a polyhedral fan.

A relation of CEGM amplitudes with Grassmannian cluster algebras \cite{Henke:2019hve,Drummond:2020kqg,Drummond:2019qjk,Drummond:2019cxm,SpeyerW,Arkani-Hamed:2020tuz,Arkani-Hamed:2019plo,Gates:2021tnp,Henke:2021ity}, 
positroid subdivisions \cite{Lukowski:2020dpn,Early:2019eun,Early:2019zyi}, and new stringy canonical forms \cite{Arkani-Hamed:2019mrd,He:2020ray} has been outlined recently. The case $k=4$ is also especially interesting due to its connection with the symbol alphabet of $\mathcal{N}=4$ SYM \cite{Henke:2019hve,Arkani-Hamed:2019rds}.

Based on the application of metric tree arrangements for parametrizing \trg(3,n)  \cite{herrmann2009draw}, Borges and Cachazo introduced a diagrammatic description of the biadjoint amplitude $m^{(3)}_n(\mathbb{I}|\mathbb{I})$, as a sum over such arrangements instead of single Feynman diagrams \cite{Borges:2019csl}. This was then extended to the case $k=4$ by Cachazo, Gimenez and the authors \cite{Cachazo:2019xjx}: In this case the building blocks are not collections but arrays (matrices) of planar Feynman diagrams. Each entry $\mathcal{M}_{ij}$ of the matrix is a planar Feynman diagram with respect to the canonical ordering $\{1,\ldots,n \} \textbackslash \{i,j\}$. We endow the diagram $\mathcal{M}_{ij}$  with a tree metric $d^{(ij)}_{kl}$ and require that it defines a completely symmetric tensor $\pi_{ijkl}:= d^{(ij)}_{kl}$. The contribution to the biadjoint amplitude is obtained by defining the function

\begin{equation}
    \mathcal{F}(\mathcal{M}):= \frac{1}{4!}\sum_{i,j,k,l} s_{ijkl}\pi_{ijkl}\, ,
\end{equation}
where $s_{ijkl}$ are generalized kinematic invariants \cite{Cachazo:2019ngv}, namely totally symmetric tensors satisfying generalized on-shell condition $s_{ii\cdots}=0$ and momentum conservation:

\begin{equation}\label{momc}
\sum_{j,k,l} s_{ijkl} = 0\,  \qquad\, \forall i.
\end{equation}
We then  compute
\begin{equation}
   \mathcal{R}(\mathcal{M}) = \int_{\Delta} d^{3(n-5)}f_ I \times e^{\mathcal{F}(\mathcal{M})}\,,
\end{equation}
where $f_I$ are the independent internal lengths and $\Delta$ the domain where \textit{all} internal lengths are positive. See \cite{Cachazo:2019xjx} for more details.\footnote{The symbol $\mathcal{C}$ was used in \cite{Cachazo:2019xjx} to denote collections, namely $k=3$ objects. Here we will respect this notation but further use $\mathcal{M}$ to denote both arrays ($k=4$) and higher rank objects made of Feynman diagram entries.} If we define by $J(\alpha)$ the set of all planar arrays for the ordering $\alpha$, the biadjoint amplitude for two orderings is then \cite{Cachazo:2019ngv}

\begin{equation}
    m^{(4)}_n (\alpha | \beta) = \sum_{\mathcal{M} \in J(\alpha)\cap J(\beta)}   \mathcal{R}(\mathcal{M}) \,.
\end{equation}

In this follow-up note we introduce a new representation of the poles of this amplitude, for the most general case $\alpha=\beta=\mathbb{I}_n$, and also discuss the general $k$ setup. We then explain how the amplitude can be recovered from such poles when they are understood from  planar arrays of Feynman diagrams. 
This new representation corresponds to collections/arrays of \textit{degenerate} ordinary Feynman diagrams and can easily be translated to  kinematic invariants in terms of $s_{\cdots}$.
This way, we get all  the poles for $(k,n)=(3,7),(3,8),(3,9),(4,7),(4,8)$ and $(4,9)$, which agree with those obtained in \cite{He:2020ray} from the stringy canonical form construction.
Conversely, as  explained in Section \ref{sec3}, we will also explain a method to get all the new presentations of poles as degenerate  arrays  by applying \textit{hard limits} to kinematic invariants  (in the sense defined in \cite{Sepulveda:2019vrz,Cachazo:2019xjx}).  
We also provide a map for which any such array defines a ray satisfying tropical Plucker relations and hence lies inside the Tropical Grassmannian polytope $\trg(k,n)$.

In an ancillary file we provide the explicit degenerate planar arrays for the cases $(k,n)=(3,6), (3,7),(3,8),(3,9)$, $(3,10),(4,7),(4,8)$ and $(4,9)$.
Those of $(3,10)$ are obtained by translating the poles given in \cite{He:2020ray} but one in principle can also get  $(3,10)$ planar collections of Feynman diagrams first using the bootstrap method given in \cite{Cachazo:2019xjx} and them degenerate them to get all poles in their new representations.  

Given two degenerated planar rays $\mathcal{V}_1,\mathcal{V}_2$ as poles, we derive and prove a criteria to check compatibility in terms of their Feynman diagram components. We show that this criteria can be translated to the weak separation condition recently studied in the mathematical literature, see e.g. \cite{speyer2020positive,arkanihamed2020positive,Early:2019eun,Early:2020hap}. We then provide an operation of addition which is equivalent to the Minkowski sum of the corresponding tropical vectors. Using this operation we can reconstruct the full facets, e.g. the planar collections and arrays previously obtained in \cite{Borges:2019csl,Cachazo:2019xjx}. More generally we can construct a graph of compatibility relations, where poles correspond to vertices and facets correspond to maximal \textit{cliques}. This gives a realization of our polytope as a Flag Complex, as observed long ago for the original construction of $\trg(3,6)$ \cite{speyer2004tropical}. We provide a Mathematica notebook as an ancillary file to implement this algorithm for all $(k,n)$ belonging to $(3,6)-(3,10)$ and $(4,7)-(4,9)$. 


This paper is organized as follows. In Section \ref{sec2} we construct poles as one-parameter arrays of degenerate Feynman diagrams and explain how to sum them to obtain higher dimensional objects, such as the full arrays of \cite{Cachazo:2019xjx}. We focus on $k=3,4$. We then provide the compatibility criteria for general $k$ and give a Mathematica implementation of the compatibility graphs. In Section \ref{sec3} we present both a kinematic and combinatoric description of the soft and hard limits of the  planar arrays of Feynman diagrams. We use it to construct the arrays corresponding to our poles, and further prove the general implementation of Grassmannian duality conjectured in \cite{Cachazo:2019xjx}. In the Discussion we outline future directions as well as a relation with matroid subdivision of the hypersimplex $\Delta(k,n)$.


\section{From Full Arrays to Poles and Back}\label{sec2}

Here we initiate the study of higher $k$ poles from the perspective
of collections of tree diagrams. Recall that these correspond to vertices
of the dual polytope (up to certain redundancies we will review), or alternatively to facets of the  positive geometry associated to stringy canonical forms \cite{Arkani-Hamed:2019mrd}.
From the perspective of the collections we shall find that the vertices
are, in a precise sense, collections of $k=2$ poles. The full collections
of cubic diagrams, studied in \cite{Cachazo:2019xjx}, correspond to convex facets
of the $(k,n)$ polytope and can be expressed as (Minkowski) sums
of poles. We start the illustration with $k=3$ and then the general cases.

\subsection{$k=3$ Planar Collections,  Poles and Compatibility Criteria }\label{k=3poles}

A very illustrative example is the case of the bipyramidal facet appearing
in $(3,6)$ \cite{Cachazo:2019apa}. This is a $k=3$ example but the discussion
will readily extend to arbitrary $k$. The description of the bipyramid
in terms of a collection has been done in \cite{Borges:2019csl,Cachazo:2019xjx}. From there,
we recall

\begin{equation}\label{bipcol}
  \centering
    \includegraphics[width=1.0\textwidth]{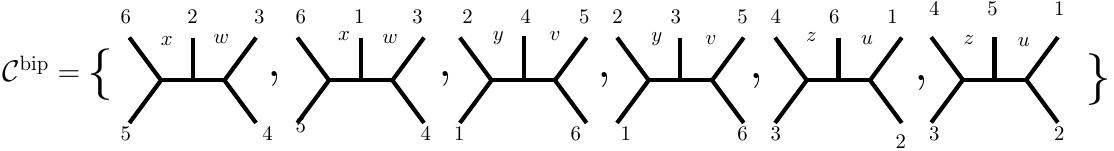}
\end{equation}
We denote each planar diagram in the collection vector by $T_i^{\textrm{bip}}$, i.e. $\mathcal{C}^{\textrm{bip}}=\{T^{\textrm{bip}}_1 \ldots T^{\textrm{bip}}_6 \}$. Note that $T_i^{\textrm{bip}}$ does not contain the $i$-th label. We have labeled by $x,y,z,u,v,w>0$ six internal distances which are 
independent solutions to compatibility conditions \eqref{eq:k3comp} below, imposed on the distances $d^{(i)}_{jk}$ from leaf $j$ to $k$ according to the graph $T_i^{\textrm{bip}}$.

\begin{figure}
  \centering
    \includegraphics[width=0.5\textwidth]{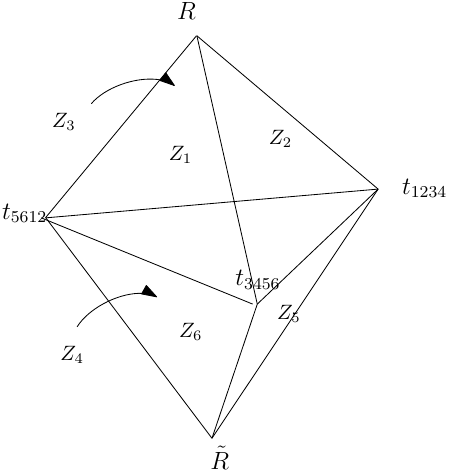}
      \caption{Bipyramid projected into three dimensions. \label{bipdr}}
\end{figure}

That this collection has the geometry of a bipyramid is seen as follows. We define its six faces by each of the allowed degeneration of metric tree arrangement. These correspond to the hyperplanes
\begin{eqnarray}
\label{Zbnd}
Z_1:  x=0 ,\qquad Z_2: y=0,\qquad\qquad\qquad\quad \,\,\,\,
Z_3: z=0\,, \qquad
\qquad
\quad\,\,\,
\\
\nonumber 
Z_4:  w=0 ,\qquad Z_5: u \equiv y-z+w=0 ,\qquad Z_6: v \equiv x-z+w=0\,,
\end{eqnarray}
in $\mathbb{R}^4$. We can project the planes into three dimensions by imposing an inhomogenous constraint, e.g. $x+y+z+w=1$, leading to the picture of Figure \ref{bipdr}. The region where all the distances in \eqref{bipcol} are strictly positive corresponds to the interior of the bipyramid. The notation for the vertices of the figure, e.g. $t_{1234},t_{3456},t_{5612},R,\tilde{R}$, will become natural in a moment. Two faces that meet at an edge correspond to two
simultaneous degenerations. For instance the edge $\{R,t_{1234}\}$
is given by setting $y=0$ and $z=0$, leading to the two-parameter collection:

\begin{equation}\label{coledge}
    \includegraphics[width=1.0\textwidth]{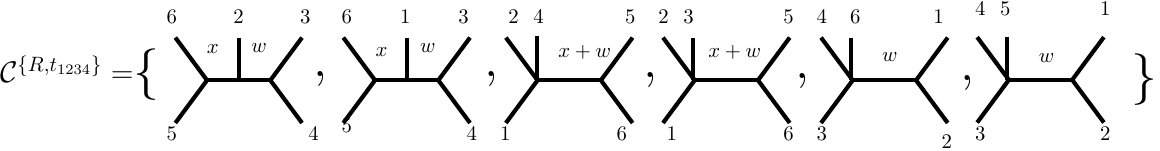}
\end{equation}

A vertex is given by three or more simultaneous degenerations. Indeed,
by performing each of the two valid degenerations of the edge,
that is $x\to0$ or $w\to0$, we arrive at the following vertices: 

\begin{equation}\label{polcol}
    \includegraphics[width=1.0\textwidth]{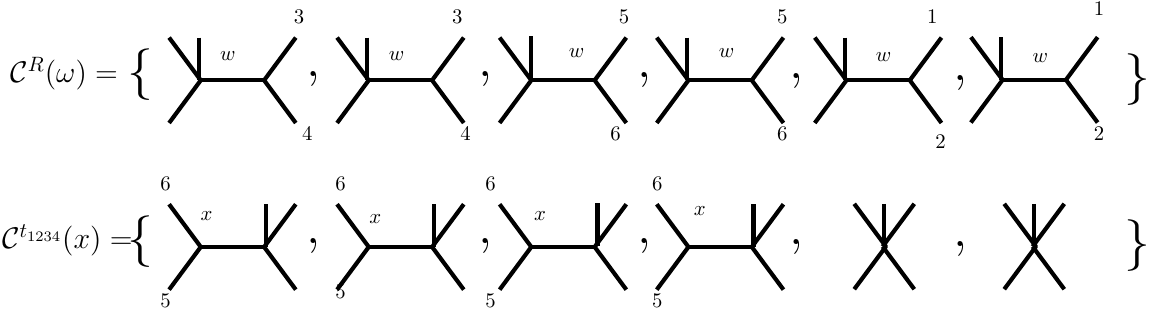}
\end{equation}

We say that two vertices are compatible if they are connected by an
edge, which can be either in the boundary or in the interior of a facet. For readers familiar with the notion of tropical hyperplanes, an edge emerges when the sum of the two independent solutions of the tropical hyperplane equations (e.g. two different vertices in \trg)  is
also a solution, see e.g. \cite{Drummond:2019qjk}.

Inspection of the facets of $(3,6)$ show that they are all convex
and hence any two vertices in a facet are compatible. In the context of  planar arrays of Feynman diagrams we will prove below that this is a general fact for all $(k,n)$. This is also motivated by the original work \cite{speyer2004tropical}, where it was argued that the polytope $(3,6)$ can be characterized as a \textit{Flag Complex}, which we can introduce as follows:

\begin{defn}\label{dff1}
The Flag Complex associated to a graph $G$ is a simplicial complex (a collection of simplices) such that each simplex is spanned by a maximal collection of pairwise compatible vertices in $G$.
\end{defn}

As a graph the bipyramid corresponds to a 4-dimensional simplex in the sense that it is spanned by the five vertices $\{R,\tilde{R},t_{1234},t_{3456},t_{5612}\}$, which are pairwise compatible, i.e. connected by an edge. However, the edge $\{ R,\tilde{R}\}$ is contained inside the bipyramid, which is equivalent to the geometrical fact that the simplex degenerates from four to three dimensions, as we explain below. The three dimensional object drawn in Figure \ref{bipdr} is what we interpret as a facet of the polytope. In general, any simplex of the Flag Complex associated to $(k,n)$ is indeed a geometrical facet, and can be embedded in $(k-1)(n-k-1)-1$ dimensions.

Following the guidelines from Tropical Geometry it is convenient to interpret each vertex as a ray
in the space of metrics embedded in $\mathbb{R}^{\binom{n}{3}}$.
Explicitly, for a compatible collection let us denote

\begin{equation}\label{eq:k3comp}
    \pi_{ijk}:=d_{ij}^{(k)}=d_{jk}^{(i)}=d_{ki}^{(j)}\,.
\end{equation}
For a vertex, $\pi_{ijk}$ depends on a single parameter $x>0$, and indeed we can write it as a ray $\pi_{ijk}\sim xV_{ijk}$.
The equivalence $\sim$ here means that we are modding out by the shift 
\begin{equation}
    \pi_{ijk}\sim\pi_{ijk}+w_{i}+w_{j}+w_{k}
\end{equation}
for arbitrary $w_i$. This is a redundancy characteristic of tropical hyperplanes, see e.g. 
\cite{herrmann2009draw}. An edge then corresponds to the Minkowski sum of two
vertices. For instance, the edge $\{R,t_{1234}\}$ corresponds to a plane in $\mathbb{R}^{\binom{n}{3}}$ given by

\begin{equation}
\pi_{ijk}^{\{R,t_{1234}\}}(x,y)\sim w\,V_{ijk}^{R}+x\,V_{ijk}^{t_{1234}}\,,\quad w,x>0.\label{eq:msumlabel}
\end{equation}

\begin{figure}
  \centering
    \includegraphics[width=0.8\textwidth]{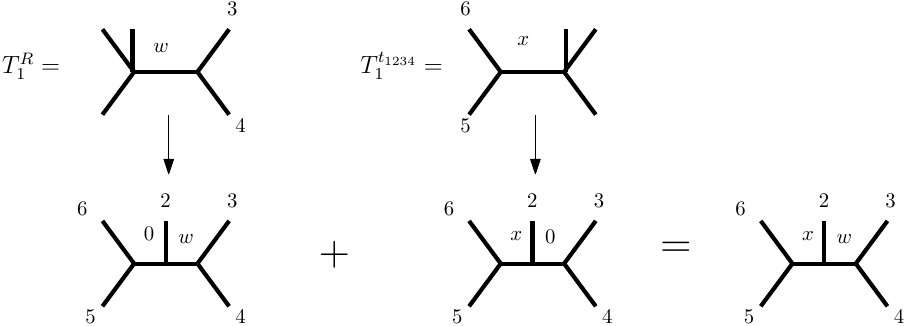}
    \caption{Addition of compatible diagrams.}\label{add}
\end{figure}

We can now justify our notation for the vertices $\{R,\tilde{R},t_{1234},t_{3456},t_{5612}\}$, namely argue for a correspondence between the vertices and kinematic poles. Indeed, the relation \eqref{eq:msumlabel} can be expressed in terms of generalized $k=3$ kinematic invariants.  Under the support of $k=3$ momentum conservation
\begin{equation}\label{momk3}
    \sum_{j,k} s_{ijk} =0\,, \qquad \forall \, i\,,
\end{equation} 
the $\sim$ symbol turns
into an equality:
\[
\mathcal{F}(\{R,t_{1234}\}):=\frac{1}{6}\sum_{ijk} s_{ijk}\pi_{ijk}^{\{R,t_{1234}\}}(x,y)=x\,R+y\,t_{1234}\,,
\]
where we used the definitions
\begin{eqnarray}
t_{1234}:=& s_{123}+s_{124}+s_{134}+s_{234}\,,\\
R=R_{12,34,56}:=&t_{1234}+s_{345}+s_{346}\,.
\end{eqnarray}
Assuming that we know the collections \eqref{polcol}, we can argue that (\ref{eq:msumlabel})
indeed defines an edge of the polytope, meaning it is associated with a compatible collection of Feynman diagrams (not necessarily cubic). In other words, we can recover \eqref{coledge}. In fact, as $\pi_{ijk}^{\{R,t_{1234}\}}(x,y)$ in \eqref{eq:msumlabel} leads to a fully symmetric
tensor $d_{ij}^{(k)}(x,y)$ by construction, it is only needed to
show that this tensor is the metric of a certain collection.
This follows from the fact that the $i$-th elements of $\mathcal{C}^{R}$ and $\mathcal{C}^{t_{1234}}$, say $T_{i}^{R}$
and $T_{i}^{t_{1234}}$ respectively, have compatible poles in the sense of $k=2$ Feynman diagrams. For instance, $T_{1}^{R}$ has the pole $s_{34}$
and $T_{1}^{t_{1234}}$ has the pole $s_{56}$, which are compatible.
In fact, the quartic vertex in $T_{1}^{R}$ can be blown up to accommodate
for the pole $s_{56}$ and vice versa. As the two diagrams now have
the same topology their metrics can be added, thus showing that $d^{(1)}_{ij}$ is indeed associated to a $k=2$ Feynman diagram. We depict this in Figure
\ref{add}. Repeating this argument for all the components $T_i$ we obtain the following \footnote{The `only if' part is easily seen to follow from a known fact of $k=2$ (applied to each of the components $T_i$): If the sum of two vertices $d_{ij}(x,y) \sim x V^X_{ij} + y V^Y_{ij}$ corresponds to a Feynman diagram (i.e. to a tropical hyperplane or a line in $\trg(2,n)$) then the two vertices $V^X,V^Y$ are compatible as $k=2$ poles.}

\begin{thm}\label{th1}
Two one-parameter collections $\mathcal{C}^{X}$ and $\mathcal{C}^{Y}$ are compatible if and only if their respective $k=2$ components $T_{i}^X$ and $T_{i}^Y$ are compatible for all $i$. 
\end{thm}

Now, an important property of $k=2$ poles is the following: Given a set of poles that are pairwise compatible, e.g. $\{s_{12},s_{123},s_{56}\}$, then such poles are compatible \textit{simultaneously}, meaning that there exists a $k=2$ Feynman diagram that includes them all.\footnote{This is seen as follows: Two compatible $k=2$ poles can be written as $s_a$ and $s_A$, where $a\subset A$ as planar subsets. Another pole that is compatible with $s_A$ can be written as $s_b$ where $b\subset A$. As $s_a$ and $s_b$ are compatible we have either $b\subset a$,$a\subset b$ or $a\cap b = \emptyset$. In all cases there exists a Feynman diagram with all three poles.} Considering this together with Theorem \ref{th1}, we can further state the following

\begin{cor}
A set of collections $\{C^{X_i}\}$ which is pairwise compatible is also simultaneously compatible.
\end{cor}

For collections, simultaneous compatibility means that the Minkowski sum of the corresponding vertices also leads to a collection. This is best explained with our example: Using Theorem \ref{th1} we can easily check that the vertices in $\{R,\text{\ensuremath{\tilde{R}}},t_{1234},t_{3456},t_{5612}\}$ are compatible in pairs. This then implies that the Minkowski sum
\begin{equation}
\pi_{ijk}^{{\rm bipyramid}}(\alpha^{i})\sim\alpha^{1}V_{ijk}^{R}+\alpha^{2}V_{ijk}^{\tilde{R}}+\alpha^{3}V_{ijk}^{t_{1234}}+\alpha^{4}V_{ijk}^{t_{3456}}+\alpha^{5}V_{ijk}^{t_{5612}}\,,\quad\alpha^{i}>0\,.\label{eq:verbip}
\end{equation}
will also be associated to an arrangement of Feynman diagrams. As we anticipated, this has the topology of a 4-dimensional simplex (in a projective sense). However, it degenerates to three dimensions due to the identity
\begin{equation}
V_{ijk}^{R}+V_{ijk}^{\tilde{R}}\sim V_{ijk}^{t_{1234}}+V_{ijk}^{t_{3456}}+V_{ijk}^{t_{5612}}
\end{equation}
or, in terms of generalized kinematic invariants,
\begin{equation}
R+\tilde{R}=t_{1234}+t_{3456}+t_{5612}\,.
\end{equation}
This implies that the Minkowski sum \eqref{eq:verbip} can be rewritten as

\begin{equation}
\pi_{ijk}^{{\rm bipyramid}}(\alpha^{i})\sim(\alpha^{1}+\alpha^{5})V_{ijk}^{R}+(\alpha^{3}+\alpha^{2})V_{ijk}^{t_{1234}}+(\alpha^{4}+\alpha^{2})V_{ijk}^{t_{3456}}+(\alpha^{5}+\alpha^{2})(V_{ijk}^{t_{5612}}-V_{ijk}^{R})\label{eq:bip}
\end{equation}
which indeed lives in three dimensions and spans the bipyramid facet. We can obtain
a more familiar parametrization of the facet, as given in \cite{Borges:2019csl}. First, let us project again the Minkowski sum
into kinematic invariants, i.e.
\begin{align*}
\mathcal{F}({\rm bipyramid}) & :=\sum_{ijk}{\bf s}_{ijk}\pi_{ijk}^{{\rm bipyramid}}(\alpha^{i})\\
 & =\underbrace{(\alpha^{1}+\alpha^{5})}_{w}R+\underbrace{(\alpha^{3}+\alpha^{2})}_{x}t_{1234}+\underbrace{(\alpha^{4}+\alpha^{2})}_{y}t_{3456}+\underbrace{(\alpha^{5}+\alpha^{2})}_{z}(t_{5612}-R)\,,
\end{align*}
where we have introduced new variables: Because $\alpha^{i}>0$
we clearly have $x,y,z,w>0$. Moreover, one can check that two independent
linear combinations of the new variables,
\begin{align}
y+w-z & =\alpha^{1}+\alpha^{4}\\
x+w-z & =\alpha^{1}+\alpha^{3}
\end{align}
are also positive. These are nothing but the $u,v>0$ conditions that
we have started with, and recover our description of the collection $\mathcal{C}^{\textrm{bip}}$, eq.  \eqref{bipcol}, whereas $\mathcal{F}({\rm bipyramid})$ agrees with that given in \cite{Borges:2019csl}. Our compatibility criteria
thus allowed us to translate back a description in terms of the vertices
(\ref{eq:verbip}) (a Minkowski sum) to a description in terms of
the full collection of cubic diagrams.

\subsection{Planar Arrays and $k>3$ Poles}\label{k>3poles}

The previous approach can be extended to the case $k>3$. Planar arrays of Feynman diagrams for $k=4$ and higher were defined in \cite{Cachazo:2019xjx} as rank $k-2$ objects, and involve the natural generalization  of the compatibility condition \eqref{eq:k3comp}.

Using the \textit{second bootstrap} approach introduced there, one can obtain matrices of planar cubic Feynman diagrams for $(k,n)=(4,7)$ starting from collections of $(k,n)=(3,6)$. A few interesting features arise for poles of $k=4$. Again, let us examine a particular planar array of $(4,7)$ in order to illustrate the construction of the poles.

The following array can be obtained from a $(3,7)$ collection via the duality procedure of \cite{Cachazo:2019xjx}, which we review in the next section. It is given by the symmetric matrix:

\begin{equation}\label{bipcol2}
  \centering
    \includegraphics[width=0.7\textwidth]{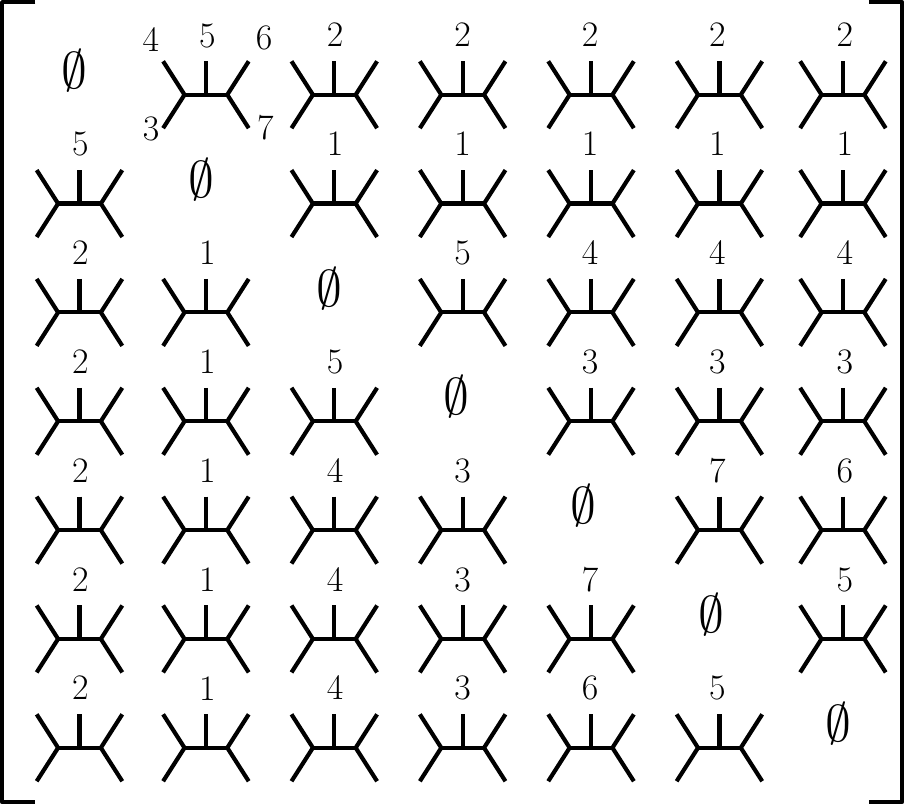}
\end{equation}
(note that the omitted labels in the Feynamn diagram entries can be deduced by their planarity). Recall that in this case the metric compatibility condition reads 
\begin{equation}\label{47comp}
    d^{(ij)}_{kl}\text{ is permutation invariant in } \{i,j,k,l\}\,,
\end{equation}
where $d^{(ij)}_{kl}$ corresponds to the distance between leaves $k$ and $l$ acoording to the diagram $\mathcal{M}^{ij}$ of the matrix. In this case the kinematic function reads
\begin{align}\label{FTM}
    \mathcal{F}(\mathcal{M}):=& \sum_{i<j<k<l} s_{ijkl}    d^{(ij)}_{kl} \nonumber \\
    =& \,\,z \,(R_{3456712} - R_{5671234} +   t_{1234} - t_{12345}) \nonumber \\ & - w\,t_{12345} + p\,(-t_{1234} + t_{12345}) - q\,t_{34567} + 
 y\,(-R_{3456712} + t_{12345} + 
    t_{34567})\nonumber  \\
   & +x\,(R_{5671234} - t_{1234} - t_{34567} -  t_{56712})\,,
\end{align}
where
\begin{eqnarray}
    R_{1234567}&=&t_{12345}+s_{1237}+s_{1236} \\
    t_{12345}&=& \sum_{J\subset \{1,2,3,4,5\}} s_J
\end{eqnarray}
together with the corresponding relabelings. Note that for a given column $i$ compatibility implies 

\begin{equation}
    d^{(ij)}_{kl}=d^{(il)}_{jk}=d^{(ik)}_{lj}\,,
\end{equation}
hence such column must be a collection, i.e. corresponds to $k=3$. For $(3,6)$ such collections can only be bipyramids (as the example of the previous section) or simplices (if they have four boundaries) \cite{speyer2004tropical}. In fact, for our example we can write the array $\mathcal{M}^{ij}$ as a ``collection of collections'' as follows:

\begin{equation}\label{47cc}
    \includegraphics[width=1.0\textwidth]{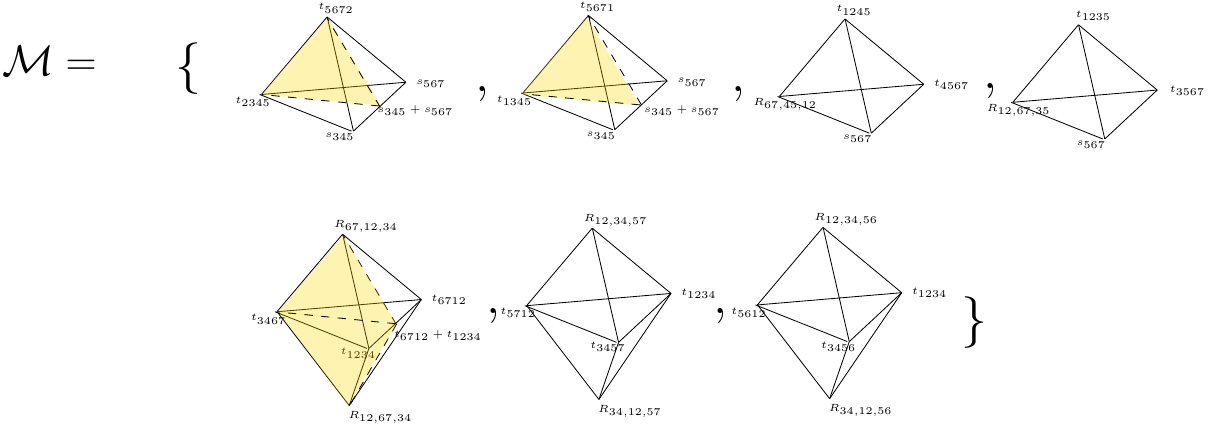}\,.
\end{equation}

Each entry represents a $(3,6)$ collection, namely a vector of cubic diagrams. They are labeled by kinematic poles as in the previous section, where collection $C^{(i)}$ contains labels $\{1,\ldots,7\} \textbackslash \{i\}$. Besides the standard degenerations (boundaries) of each $C^{(i)}$, we have depicted some internal boundaries in yellow. These arise from the external boundaries of another collection, say $C^{(j)}$, through the compatibility condition \eqref{47comp}. For instance, translating the bipyramid boundaries \eqref{Zbnd} to the new variables we used in \eqref{FTM} we see that collection $C^{(5)}$ has the following six boundaries: 

\begin{align}\label{c5planes}
  C^{(5)}:
  \,\,\, x=0\,,\quad  y=0\,,\quad  w=0\,,\quad q=0
  \,, \nonumber \\
    w + x - y=0\,,\quad  q + x - y=0 \,,
\end{align}
which depend only on four variables $\{x,y,w,q\}$ instead of six, as expected for a bipyramid living in three dimensions. Now, we further consider the following (external) degenerations of collections $C^{(1)}$ and $C^{(3)}$,
\begin{align*}
    C^{(1)}:\, & -p + w - y + z=0\,, \quad  z=0 \nonumber \\
     C^{(3)}:\,& \qquad\,  p=0 \,, 
\end{align*}
which altogether induce the plane $w=y$ as a new degeneration, depicted by the internal yellow plane bisecting the bipyramid $C^{(5)}$ in \eqref{47cc}. The intersection of this plane with its external faces $q=q + x - y=0$ in \eqref{c5planes} induces a new ray in $(3,6)$, obtained as the midpoint of the vertices $t_{1234}$ and $t_{6712}$, labelled as $t_{1234}+t_{6712}$ in \eqref{47cc}. In the other collections $C^{(i)}$, the induced (3,6) rays can be further labeled in the same way and lead to the (4,7) ray we denote $W$:

\begin{equation}\label{vw}
    \mathcal{M}^W  = \{s_{345}+s_{567}, s_{345}+s_{567},R_{67,45,12},R_{12,67,35},t_{6712}+t_{1234},R_{12,34,57},R_{12,34,56}\}\,,
\end{equation}
while the kinematic function \eqref{FTM} becomes
\begin{equation}\label{vwk}
    \mathcal{F}(\mathcal{M}^W) = x W_{1234567} := x \left(\sum_a \,(s_{a567}+s_{a345})+s_{3467}\right )\,.
\end{equation}
In \eqref{vw} the sum of vertices must be understood in the sense of the previous section. That is, we consider the line $x V_{abc}^{t_{6712}}+ y V_{abc}^{t_{1234}}$, which belongs to the $k=3$ polytope since $t_{6712}$ and $t_{1234}$ are compatible (in fact, they appear together in a bipyramid). The vertex $t_{6712}+t_{1234}$ corresponds to its midpoint $x=y$. Addition of (collections of) Feynman diagrams is done as in Fig. \ref{add}, for instance: 

\begin{equation}\label{polcol2}
    \includegraphics[width=1.0\textwidth]{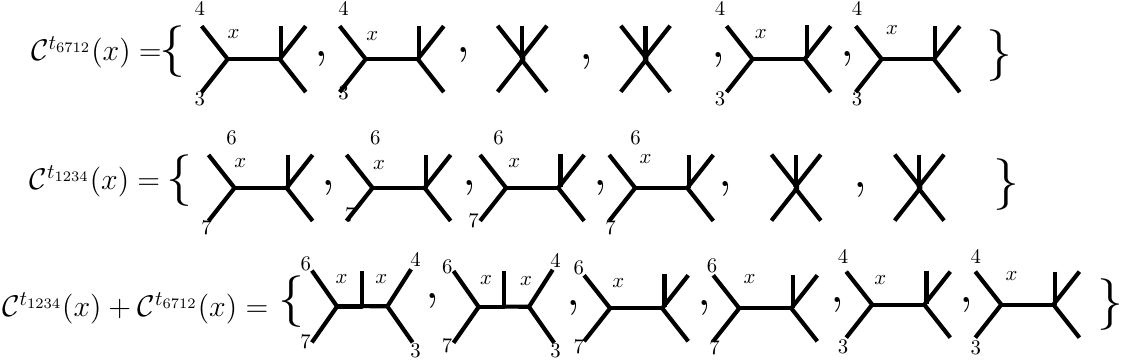}
\end{equation}
which can also be written more compactly as
\begin{eqnarray}\label{compa}
\mathcal{C}^{t_{6712}} &=& \{s_{43},s_{43},0,0,s_{43},s_{43}\} \,,\nonumber \\
\mathcal{C}^{t_{6712}} &=& \{s_{67},s_{67},s_{67},s_{67},0,0\}\,,
\nonumber \\
    \mathcal{C}^{t_{1234}}+\mathcal{C}^{t_{6712}} &=& \{s_{67}+s_{43},s_{67}+s_{43},s_{67},s_{67},s_{43},s_{43}\}\,.
\end{eqnarray}
(Note that $s_{43}$ and $s_{67}$ are compatible, thus their sum also belongs to the $k=2$ polytope.) Hence each of the entries of \eqref{vw} can be represented as a column, and $\mathcal{M}^W$ can be written as a $7\times 7$ matrix with a single internal distance parameter $x$. This is precisely our original array of cubic diagrams \eqref{bipcol2} after the degenerations have been imposed.

We have learned that the compatibility condition \eqref{47comp} can lead to particular boundary structures as in \eqref{vw}. Because of this the vertex $W$ of (4,7) is not only decomposed in terms of vertices of (3,6) but also certain internal rays (midpoints) of the (3,6) polytope. It would be interesting to classify the kind of internal rays that can appear in this decomposition.

\subsubsection*
{Compatibility criteria for general $k$}

We now present the criteria for compatibility of poles, which 
easily extends from the case $k=3$ 
in the previous part of this section to general $k$. A vertex of the $(k,n)$ polytope, realized as a ray in the space of metrics, will be determined by a completely symmetric array of rank $k-2$. From now on we will denote such array as $\mathcal{V}^{(k,n)}_{i_1\ldots i_{k-2}}$,\footnote{Here $\mathcal{V}$ denotes a generic vertex. For a particular vertex labeled by $X$ we may use $\mathcal{M}^X$ as in the previous section.} e.g. a vector for $k=3$ and a matrix for $k=4$, where each component is a planar Feynman diagram.  

 Such array can be organized by `columns' we define by $\mathcal{T}^{(i)}_{i_2\ldots i_{k-2}} :=\mathcal{V}^{(k,n)}_{i i_2\ldots i_{k-2}} $, so we write:
\begin{equation}\label{gnr}
\mathcal{\mathcal{V}}^{(k,n)}=[\mathcal{T}^{(1)},\ldots,\mathcal{T}^{(n)}]
\end{equation}
where $\mathcal{T}^{(i)}$ are rays in $(k-1,n-1)$, which can also be written as arrays of Feynman diagrams. In the previous example this corresponds to our decomposition \eqref{vw}, where $\mathcal{V}^{(4,7)}=\mathcal{M}^W$ and $\mathcal{T}^{(i)}$ are $k=2$ rays. 

In the $k=2$ polytope two rays $d_{ij}(x)$ and $d'_{ij}(y)$ are compatible (their sum corresponds to a Feynman diagram) if and only if their respective kinematic poles are compatible. For general $k$ this means that the components $d^{i_1\ldots i_{k-2}}_{ij}(x)$ and $d'^{i_1\ldots i_{k-2}}_{ij}(y)$ are compatible if and only if the diagrams $\mathcal{V}^{(k,n)}_{i_1\ldots i_{k-2}}$ and $\mathcal{V}'^{(k,n)}_{i_1\ldots i_{k-2}}$ are compatible in the $k=2$ sense. Furthermore, two rays represented by arrays $\mathcal{V}^{(k,n)}$ and $\mathcal{V}'^{(k,n)}$ are compatible when all their diagrams are. This implies the following extension of theorem \ref{th1}

\begin{thm}
\label{tm2}
Two vertex arrays $\mathcal{\mathcal{V}}^{(k,n)}$ and $\tilde{\mathcal{V}}^{(k,n)}$
are compatible if and only if their components $\mathcal{T}^{(i)}$ and $\tilde{\mathcal{T}}^{(i)}$
are compatible as rays of $(k-1,n-1)$, for all $i=1,\ldots,n$.
\end{thm}

Assuming we know all possible poles, this provides an efficient criteria for checking their compatibility and, repeating the Minkowski sum construction of the previous section, constructing the facets of the polytope. In fact, this criteria has already appeared in the math literature in the context of \textit{boundary maps} of the hypersimplex $\Delta_{k,n}$, see e.g. \cite{speyer2020positive,arkanihamed2020positive,Early:2019eun}; \footnote{We thank N. Early for pointing this out.} we will establish the precise equivalence in section \ref{sec:disbnd}. 

\begin{cor}
Two vertex arrays $\mathcal{\mathcal{V}}^{(k,n)}$ and $\tilde{\mathcal{V}}^{(k,n)}$
are compatible if and only if their entries $\mathcal{T}^{(i_1,i_2,\cdots,i_{k-2})}$ 
and  $\tilde{\mathcal{T}}^{(i_1,i_2,\cdots,i_{k-2})}$
are compatible for all $\{i_1,i_2,\cdots,i_{k-2}\} \subset \{1,2,\cdots,n\}$.
\end{cor}

The proof is straightforward.

The full list of poles can be obtained by degenerations of just a few known facets of the polytope, given in our previous work \cite{Cachazo:2019xjx}, or more simply from singularities in the moduli space through the $\mathbb{CP}^{k-1}$ scattering equations of \cite{Cachazo:2019apa}, as recently done in \cite{He:2020ray}. Once the full list of poles is known as combination of kinematic invariants, we can easily translate to arrays using the procedure described in the next section. In an attached notebook, we construct the full facets for $k=3$ and $k=4$ starting from the kinematic poles and provide a simple implementation of Theorems \ref{th1} and \ref{tm2} to check the compatibility relations of any pair of poles.

\subsection{Summary of the Results}

We list the number of poles and the number of pairs of compatible poles for  $(3,6)-(3,9)$ and for $(4,7)-(4,9)$  in Table \ref{k3matrices} and Table \ref{k4matrices} respectively. What's more, we recover all the planar arrays of Feynman diagrams using the compatibility rules, whose number is also 
summarized in the tables. 
Using the $(3,10)$ poles given in \cite{He:2020ray}, we also predict that  the number of $(3,10)$ planar collections of Feynman diagrams should be $11\,187\,660$. 
We see the number of poles is much less than that of planar GFDs. 
The minimal number of poles for a $(k,n)$ planar arrays of Feynman diagrams is $(k-1)(n-k-1)$ and we list the maximal number in the tables.
We see that a (4,9) planar matrix of Feynman diagrams could have 461 poles, much larger than the minimal number 12.

\begin{table}[!htb]  
  \centering
  \begin{tabular}{ |p{4.5cm}|p{1.3cm}|p{1.3cm}|p{1.7cm}|p{1.7cm}|p{1.9cm}|}
    \cline{2-6} 
    \multicolumn{1}{c|}{} &\hspace{2mm}\textbf{(3,\,6)} &\hspace{2mm}\textbf{(3,\,7)}
    &\hspace{4mm}\textbf{(3,\,8)}
    &\hspace{4mm}\textbf{(3,\,9)}
    &\hspace{4mm}\textbf{(3,\,10)}
    \\
 \hline
 {\small {\# of poles}} & \hspace{4mm} 16 & \hspace{4mm} 42 & \hspace{5mm}
 120 & \hspace{4mm} 
471
  & \hspace{4mm} 
3140 
 \\
\hline
 {\small {\# of compatible pole pairs}} & \hspace{4mm} 68 & \hspace{3mm} 434 & \hspace{4mm}
2768 & \hspace{3mm} 
26\,949
  & \hspace{2mm}
577\,485
 \\
  \hline
  {\small {\# of planar collections of FDs }} & \hspace{4mm} 48 & \hspace{3mm} 693 & \hspace{2mm}
 13\, 612 & \hspace{2mm} 
 346\, 710
  & \hspace{0mm} 
11\,187\,660 
  \\
  \hline
{\small mimimal \# of poles each }& \hspace{5mm} 4 & \hspace{5mm} 6 & \hspace{7mm} 8 & \hspace{6mm} 10 &
\hspace{6mm} 12
\\
  \hline
{\small maximal \# of poles each }& \hspace{5mm} 5 & \hspace{5mm} 9 & \hspace{7mm}17 & \hspace{6mm} 46 &
\hspace{5mm} 123
\\
  \hline
  \end{tabular}
  \caption{Number of poles, pairs of compatible poles,  planar collections of Feynman diagrams and  minimal vs. maximal number of poles in a single planar collection of Feynman diagrams for $k=3$.}
  \label{k3matrices}
  \end{table}

\begin{table}[!htb]  
  \centering
  \begin{tabular}{ |p{4.5cm}|p{2.0cm}|p{2.0cm}|p{2.0cm}|}
    \cline{2-4} 
    \multicolumn{1}{c|}{}  &\hspace{5mm}\textbf{(4,\,7)}
    &\hspace{5mm}\textbf{(4,\,8)}
    &\hspace{5mm}\textbf{(4,\,9)}
    \\ 
 \hline
 {\small {\# of poles}}
 & \hspace{7mm} 42 & \hspace{6mm} 360 & \hspace{4mm} 19\,395
 \\
 \hline
  {\small {\# of compatible pole pairs }}
 & \hspace{6mm} 434 & \hspace{4mm} 16\,128 & \hspace{2mm} 8\,833\,230
 \\
 \hline
 {\small {\# of planar matrices of FDs }}
 & \hspace{6mm} 693 & \hspace{4mm} 90\,608 & \hspace{1mm} 30\,659\,424
     \\
  \hline
{\small minimal \# of poles each }
 & \hspace{8.5mm} 6 & \hspace{8mm} 9 & \hspace{6mm} 12
   \\
  \hline
{\small maximal \# of poles each }
 & \hspace{8.5mm} 9 & \hspace{7mm} 49 & \hspace{5mm} 461
 \\
  \hline
  \end{tabular}
  \caption{Number of poles, pairs of compatible poles,  planar matrices of Feynman diagrams and  minimal vs. maximal number of poles in a single planar matrix of Feynman diagrams for $k=4$.}
  \label{k4matrices}
  \end{table}


\section{Soft/Hard Limits and Duality}\label{sec3}

In \cite{Sepulveda:2019vrz} a kinematic hard limit has been introduced, based on the Grassmannian duality of the generalized amplitude $m^{(k)}_n$ and its soft limit. Up to momentum conservation ambiguities, we can define it as (take e.g. particle label 1)

\begin{equation}
s_{1\ldots}\rightarrow \tau \hat{s}_{1\ldots }\,,\qquad  \textrm{ with  } \tau \to \infty \,.
\end{equation}  

In this section we give a combinatoric description of such soft and hard kinematic limits in terms of the arrays. In fact, this provides a method for constructing the array of a pole only from the knowledge of the corresponding kinematic invariant. 

Very nicely, this will also provide a geometric interpretation to the \textit{combinatorial} soft and hard limits recently discussed in \cite{Cachazo:2019xjx} for facets, e.g. full collections and arrays. At the same time, it leads to a proof of the combinatorial duality proposed there for such facets.

Let us use $k=4$ as an example to introduce the connection between kinematics and combinatorics. A pole is described in term of generalized kinematic invariants by the function

\begin{equation}
    \mathcal{F}(\mathcal{V})=\frac{1}{4!}\sum s_{ijkl} d_{ij}^{(kl)}(x)  = \frac{x}{4!}\sum s_{ijkl} V_{ijkl}\,.
\end{equation}

Let us first consider the soft limit on label 1, i.e. $s_{1\ldots}\rightarrow \tau \hat{s}_{1\ldots }$ extracting the leading order as $\tau \to 0$. The pole then becomes
\begin{equation}
    \mathcal{F}(\mathcal{V})\rightarrow     \mathcal{F}^{\rm soft}_1 (\mathcal{V}) :=  \frac{1}{4!}\sum_{i,j,l,m\neq 1} s_{ijlm} \tilde{d}_{ij}^{(lm)}(x)  =\mathcal{F}(\tilde{\mathcal{V}}_1)
\end{equation}
where $s_{ijlm}$ with $i,j,l,m\neq 1$ correspond to hard kinematic invariants, they satisfy generalized momentum conservation \eqref{momc} for $n-1$ labels. Also, $\tilde{d}_{ij}^{(lm)}(x) $ corresponds to the metric of $\tilde{\mathcal{V}}_1^{lm}$: The latter is obtained by removing column 1 and row 1 in the matrix of Feynman diagrams $\mathcal{V}^{lm}$. Furthermore, the restriction $i,j\neq 1$ in $\tilde{d}_{ij}^{(lm)}(x) $ means particle $1$ can be removed in the Feynman diagrams of $\tilde{\mathcal{V}}^{lm}_1$. The object $\tilde{\mathcal{V}}^{lm}_1$ is also a one-parameter array for $k=4$, i.e. corresponds to $(4,n-1)$. It is precisely the \textit{combinatorial soft limit} of $\mathcal{V}^{lm}$ in the sense of \cite{Cachazo:2019xjx}.

The hard limit proceeds in a similiar fashion. On label 1, take $s_{1\ldots}\rightarrow \tau \hat{s}_{1\ldots }$ and extract the leading terms as $\tau \to \infty$. The result can be written as
\begin{equation}\label{hrdef}
    \mathcal{F}(\mathcal{V})\rightarrow     \mathcal{F}^{\rm hard}_1 (\mathcal{V}) =  \frac{1}{3!}\sum_{j,k,l\neq 1} \hat{s}_{jkl} \hat{d}_{ij}^{(l)}(x)  =\mathcal{F}(\mathcal{T}^{(1)})
\end{equation}
where $\hat{d}_{ij}^{(l)} =d^{(1l)}_{ij}$ and $\hat{s}_{jkl}:=s_{1jkl}$ are now interpreted as $k=3$ kinematic invariants since they satisfy $\sum_{kl} \hat{s}_{jkl}=0$. Thus, the kinematic hard limit of the collection $\mathcal{V}^{ij}$ can be obtained from its column $\mathcal{T}^{(1)}$, and corresponds to the $k=3$ collection defined as $\mathcal{C}^i$:= $(\mathcal{T}^{(1)})^i=\mathcal{V}^{1i}$. As discussed, $\mathcal{T}^{(1)}$ gives an  array of $(3,n-1)$, and indeed turns out to be the \textit{combinatorial} hard limit of $\mathcal{V}^{ij}$ in the sense of \cite{Cachazo:2019xjx}.

This has a direct application for constructing an array given certain kinematics. Take for instance the pole $W$ in $(4,7)$, eq. \eqref{vwk},
\begin{equation}
    W_{1234567} := \sum_a \,(s_{a567}+s_{a345})+s_{3467}\,.
\end{equation}
Under the hard limit in e.g. particle 7,
\begin{equation}
    W_{1234567} \rightarrow  \sum_a \,\hat{s}_{a56}+\hat{s}_{345}+\hat{s}_{346}=R_{12,34,56}\,.
\end{equation}
Thus, from \eqref{hrdef}, $R_{12,34,56}$ corresponds to the valuation of the column $\mathcal{T}^{(7)}$ in the array of $W$, i.e. $\mathcal{F}(\mathcal{T}^{(7)})$, and can be used interchangeably. Indeed, by applying the hard limits in all the labels one recovers the array
\begin{equation}
   W_{1234567} \rightarrow \{s_{345}+s_{567}, s_{345}+s_{567},R_{67,45,12},R_{12,67,35},t_{6712}+t_{1234},R_{12,34,57},R_{12,34,56}\}
\end{equation}
which justifies the notation \eqref{vw}. Of course, each of the elements here is indeed a column, which can be constructed by applying yet another hard limit, e.g.
\begin{equation}
    t_{1234}+t_{6712} \rightarrow \{s_{67}+s_{43},s_{67}+s_{43},s_{67},s_{67},s_{43},s_{43}\}
\end{equation}
which is \eqref{compa}. Thus, the matrix associated to $ W_{1234567}$ is obtained by a consecutive double hard limit. The fact that the matrix $(\mathcal{M}^W)_{ij}$ obtained so is symmetric corresponds to the statement that the two hard limits commute.

For a general vertex of the $(k,n)$ polytope, of the form \eqref{gnr}, we have
\begin{align}\label{fvc}
      \mathcal{F}^{\rm soft}_1 (\mathcal{V}) =& \mathcal{F}(\tilde{\mathcal{V}}_1) \nonumber \\
       \mathcal{F}^{\rm hard}_1 (\mathcal{V}) =& \mathcal{F}(\mathcal{T}^{(1)}) 
\end{align}
where $\mathcal{T}^{(1)}$ is a vertex of $(k-1,n-1)$ and $\tilde{\mathcal{V}}_1$ is a vertex of $(k,n-1)$:

\begin{equation}
    \tilde{\mathcal{V}}_1:=[\mathcal{T}^{(2)},\ldots,\mathcal{T}^{(n)}]_{\text {(1 removed)}}
\end{equation}

This provides a kinematic interpretation of the combinatorial hard ($k$-reducing) and soft ($k$-preserving) operations for general $k,n$. Also, given $\mathcal{F}(\mathcal{V})$, the array $\mathcal{V}_{i_1\ldots i_{k-2}}$ can be constructed by applying $k-2$ consecutive hard limits in labels $i_1,\ldots, i_{k-2}$ and identifying the resulting $k=2$ Mandelstam with a Feynman diagram as in \eqref{polcol2} and \eqref{compa}.

\subsection{Duality}

It is known that the $(k,n)$ polytope admits a dual description as a $(n-k,n)$ polytope, induced by Grassmanniann duality $G(k,n)\sim G(n-k,n)$. In the moduli space this identification was shown to imply the relation $m^{(k)}_n=m^{(n-k)}_n$ \cite{Cachazo:2019apa,Sepulveda:2019vrz}. In the context of the polytope (i.e. kinematic space) the identification is true for facets, edges, vertices, etc. Indeed, a duality for  planar arrays of  Feynman diagrams was conjectured in \cite{Cachazo:2019xjx} and relates arrays in $(k,n)$ and $(n-k,n)$. It is such that

\begin{equation}\label{fmd}
\mathcal{F}(\mathcal{M}^{(k,n)}) =\mathcal{F}(\mathcal{M}^{*(n-k,n)})\,, 
\end{equation}
under appropriate relabelings. Furthermore, both collections $\mathcal{M}^{(k,n)}$ and $\mathcal{M}^{*(n-k,n)}$ have the same boundaries arising as degenerations and hence lead to the same contribution to the biadjoint amplitude $m^{(k)}_n$.

We now describe and provide a proof of the duality. Let us first consider the case of vertices and then promote it to facets via the sum procedure of the previous section. Two vertices in $(k,n)$ and $(n-k,n)$ are defined as duals when they satisfy 

\begin{equation}\label{fvd}
    \mathcal{F}(\mathcal{V}^{(k,n)}) =\mathcal{F}(\mathcal{V}^{*(n-k,n)}) \,,
    \end{equation}
    i.e. they are kinematically the same.

Of course, the previous definition requires to relabel the kinematic invariants. Focusing on the case $(4,7)\sim (3,7)$ to illustrate this, the relabeling is $s_{abcd}\sim s_{efg}$ where $\{a,b,c,d,e,f,g\}=\{1,\ldots,7\}$. Suppose now $a=1$. The hard limit in label 1 of $s_{1bcd}$ 

\begin{equation}
    s_{1bcd}\to  \hat{s}_{bcd}\,,
\end{equation} 
while under the soft limit 
\begin{equation}
    s_{efg} \to s_{efg}\,.
\end{equation}
But $\{b,c,d,e,f,g\}=\{2,\ldots,7\}$ and hence $\hat{s}_{bcd} \sim s_{efg}$ under the duality $(3,6)\sim (3,6)$. This can be repeated while replacing $s_{1bcd}$ by any linear combination of kinematic invariants, in particular by the one given by $\mathcal{F}(\mathcal{V})$. The conclusion can be nicely depicted by the diagram: 
\begin{equation}\label{dl}
    \includegraphics[width=0.5\textwidth]{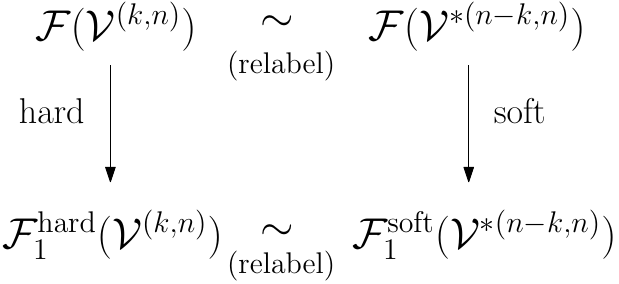}
\end{equation}
But from \eqref{fvc} we conclude that $ \mathcal{F}(\mathcal{T}^{(1)})\sim \mathcal{F}(\tilde{\mathcal{V}}^{*}_1)$. That is, the $(k-1,n-1)$ array $\mathcal{T}^{(1)}$, which is the first column of $\mathcal{V}$, is dual to the $(n-k,n-1)$ array $\tilde{\mathcal{V}}^{*}_1$, which corresponds to $\mathcal{V}^{*}$ with the first components removed. Repeating the steps for all other labels this proves the following:

\begin{thm}\label{dualvv}
Let $\mathcal{V}^{*(n-k,n)}$ be the dual ray to $\mathcal{V}^{(k,n)}=[\mathcal{T}^{(1)},\ldots, \mathcal{T}^{(n)}]$, that is $\mathcal{F}(\mathcal{V})=\mathcal{F}(\mathcal{V}^*)$ under appropiate relabelings. Then the hard limit $\mathcal{T}^{(i)}$ is dual to the soft limit $\mathcal{\tilde{V}}^*_i$ for all $i=1,\ldots,n$.
\end{thm}
Since soft and hard limits always reduce the number of labels $n$, this theorem can be iterated to check whether two given rays are duals.

This criteria was conjectured for facets in \cite{Cachazo:2019xjx}. To prove the criteria for two dual facets, say $\mathcal{M}^{(k,n)}$ and $\mathcal{M}^{*(n-k,n)}$, we resort to the construction in Section \ref{sec2}. According to the characterization \ref{dff1} of the polytope, facets or full collections are maximal sums of poles. Using the notation of eq. \eqref{eq:verbip} for the space of compatible metrics for an array, we can write
\begin{equation}
    \pi^{\mathcal{M}} \sim \sum_I \alpha^I V^I \,,\quad \alpha^I>0\,.
\end{equation}
or simply, using addition of compatible Feynman diagrams,
\begin{equation}\label{msum}
    \mathcal{M}_{i_1\ldots i_{k-2}} = \sum_I \alpha^I \mathcal{V}^I_{i_1\ldots i_{k-2}} \,,\quad \alpha^I>0\,.
\end{equation}
Now let us define the following object:
\begin{equation}\label{msum2}
    \mathcal{M}^*_{i_1\ldots i_{k-2}} := \sum_I \alpha^I \mathcal{V}^{*I}_{i_1\ldots i_{k-2}} \,,\quad \alpha^I>0\,.
\end{equation}
The duality relation \eqref{fmd} follows from \eqref{fvd} together with the linearity of the map $\mathcal{F}$. Of course, this definition requires that the vertices $\mathcal{V}^{*I}$ can be added. In fact, we now prove

\begin{theorem}\label{thm32}
The set $\{\mathcal{V}^{*I}\}$ is a maximally compatible collection of vertices (a clique) of $(n-k,n)$. Hence $\mathcal{M}^*$ is a facet, the dual facet of $\mathcal{M}$.
\end{theorem}

\textit{Proof:} It suffices to show that if two vertices, say $\mathcal{V}$ and $\mathcal{W}$, are compatible, so are their duals $\mathcal{V}^*$ and $\mathcal{W}^*$. This follows from induction in $n$ (the case $n=4$ being trivial): If $\mathcal{V}$ and $\mathcal{W}$ are compatible in $(k,n)$, it is easy to see that their combinatorial soft limits $\tilde{\mathcal{V}}_i$ and $\tilde{\mathcal{W}}_i$ are compatible in $(k,n-1)$, for all $i$. Then, using the induction hypothesis we find that their duals, $(\tilde{\mathcal{V}}_i)^*$ and $(\tilde{\mathcal{W}}_i)^*$ are compatible. From Theorem \ref{dualvv} these are the hard limits of $\mathcal{V}^*$ and $\mathcal{W}^*$ for all $i$. It then follows from Theorem \ref{tm2} that $\mathcal{V}^*$ and $\mathcal{W}^*$ are themselves compatible as $(k-n,n)$ arrays, which completes the induction. $\square$

Having successfully characterized the dual facet of $\mathcal{M}$ by eq. \eqref{msum2} we can now prove the extension of Theorem \ref{dualvv} for facets. For this, let us simply denote by $\tilde{\mathcal{M}}_i$ the combinatorial soft limit of $\mathcal{M}$ in particle $i$, which indeed defines a facet of the $(k,n-1)$ polytope. From \eqref{msum}, it is easy to see that the soft limit is
\begin{equation}
    \tilde{\mathcal{M}}_i = \sum_I \alpha^I \tilde{\mathcal{V}}^I_i \,.
\end{equation}
On the other hand the combinatorial hard limit of $\mathcal{M}^*$ (given by \eqref{msum2}) in particle $i$ is
\begin{equation}\label{tim}
    \mathcal{T}_{\mathcal{M}^*}^{(i)} = \sum_I \alpha^I \mathcal{T}_{\mathcal{V}^{*I}}^{(i)} \,,
\end{equation}
where $\mathcal{M}^*= [\mathcal{T}^{(1)}_{\mathcal{M}^*},\ldots, \mathcal{T}^{(n)}_{\mathcal{M}^*}]$, etc.
As each  $\tilde{\mathcal{V}}^I_i$ is dual to $\mathcal{T}_{\mathcal{V}^{*I}}^{(i)}$, we conclude that $  \mathcal{T}_{\mathcal{M}^*}^{(i)} $ as given by \eqref{tim} is the dual facet to $\tilde{\mathcal{M}}_i$. This proves the duality criteria for facets, first proposed in \cite{Cachazo:2019xjx}.

\section{Discussion}

In \cite{Cachazo:2019xjx} a combinatorial bootstrap was introduced for obtaining the collections corresponding to facets of $(k,n)$ with $k\geq 4$. In a nutshell, for $k=4$ one writes a candidate symmetric array of Feynman diagrams using as a set of columns the facets of $(3,n-1)$, i.e.
\begin{equation}\label{mdec}
\mathcal{M}=\{\mathcal{C}^1,\ldots, \mathcal{C}^n\}\, ,
\end{equation} 
Then, one checks whether the corresponding metric $d^{(ij)}_{kl}$ can be  imposed to be symmetric, in which case one has found a facet of $(4,n)$. In this work we have explored the representation \eqref{mdec} for vertices $\mathcal{V}$ of $(k,n)$. However, in section \ref{k>3poles} we have discovered that in this case the columns $\mathcal{T}^{(i)}$ (which play the role of the $C^i$ in \eqref{mdec}) are not necessarily vertices of $(k-1,n-1)$ but rather certain internal rays in the polytope. It would be interesting to clasify which kind of internal rays can appear, as a way of implementing the combinatorial boostrap more efficiently at the level of vertices. Interestingly, for all the examples explored in this work we were able check a weaker version: A compatible collection $\{\mathcal{T}^{(1)}_V,\ldots,\mathcal{T}^{(n)}_V\}$ for which all the $\mathcal{T}^{(i)}_V$ are vertices in $(k-1,n-1)$ is indeed a vertex of $(k,n)$. For instance, for $(4,8)$ we obtain 98 poles given by all possible compatible collections of poles of $(3,7)$ (also including the trivial column).

Despite these subtleties, we found there is a universal way to translate any poles in terms of generalized kinematics as degenerated arrays and vice versa, check their compatibility relations and reconstruct the full planar arrays of Feynman diagrams for any $k\geq 3$.



\subsection{Relation to Boundary Map from the $(k,n)$ Hypersymplex}\label{sec:disbnd}

The hard limit we have introduced here can be understood as a map $(k,n)\to (k{-}1,n{-}1)$ for any ray in  $\trg(k,n)$.  Recently, Early has introduced a basis of planar poles, corresponding to certain matroid subdivisions of the hypersimplex $\Delta(k,n)$ \cite{Early:2019zyi,Early:2019eun,Early:2020hap}. Then, there is a natural boundary restriction $\partial^{(j)}\Delta_{k,n} = \Delta_{k{-}1,n{-}1}$ that can be applied in order to characterize such matroid subdivisions. When acting on Early's planar basis, we will now argue that the boundary restriction agrees with our hard kinematical limit. Since the action of the boundary restriction on a generic pole is the linear extension of the action on the basis, this means that the hard limit effectively implements the boundary map in general. 

We construct the planar basis as follows: Consider the hypersimplex $\Delta_{k,n}$ defined by

\begin{equation}
x_1+\ldots+x_n = k\,, \,\, x_j \in [0,1]\,.    
\end{equation}

Let $I\subset \{1,\ldots,n\}$ be a subset of $k$ elements, $|I|=k$. The vertex $e_I\in \mathbb{R}^n$ of the hypersimplex corresponds to $x_i=1$ for $i\in I$, with all the other $x_l=0$. The boundary $\partial^{(j)}\Delta_{k,n}$ is obtained by setting $x_j=1$, i.e. it becomes the hypersimplex $\Delta_{k{-}1,n{-}1}$ given by $\sum_{i\neq j} x_i =k-1$. Note that this decreases the value of $k$ and $n$ exactly as the hard limit introduced in Section \ref{sec3}. Focusing on the boundary $x_1=1$, let us denote its vertices by $e^{(1)}_{\hat{I}}:=e_{1\hat{I}}$ where $\hat{I}\subset \{2,\ldots,n\}$ with $|\hat{I}|=k-1$.

Positroid subdivisions can be obtained from the level function $h(x):\Delta_{k,n}\to \mathbb{R}$ studied in \cite{Early:2019zyi,Early:2019eun}. This is piecewise linear in the hypersimplex and its curvature is localized on certain hyperplanes defining the subdivision. Early's basis is in correspondence with the subdivisions arising in the set of functions

\begin{equation}\label{}
    h(x-e_J)\,\,\,, J\,\textrm{  non-consecutive.}
\end{equation}

Since there are $n$ consecutive subsets $I$, the number of such functions is $\binom{n}{k}-n$, precisely the number of independent kinematics for $(k,n)$.\footnote{In this notation a set of independent kinematic invariants (which is non planar, i.e. does not entirely correspond to poles of $m^{(k)}_n(\mathbb{I}_n|\mathbb{I}_n)$) is generated by $s_J$ with $|J|=k$. Recall that we further have $n$ momentum conservation constraints, in this notation $\sum_{|\hat{I}|=k-1} s_{j\hat{I}}=0$ for all $j=1,\ldots,n$.} To obtain the explicit planar basis in kinematic space we take the linear combination (up to an overall normalization)

\begin{equation}
    \eta_J =  \sum_{|I|=k} h(e_I-e_J) s_I \,.
\end{equation}

(One can check that by virtue of momentum conservation $\eta_J=0$ for a consecutive subset $J$ \cite{Early:2019eun}). Now we show that the hard limit as constructed in Section \ref{sec3} has precisely the same effect as the  boundary map restriction $\partial^{(1)}$ of the subdivision $h(x-e_J)$. Taking the hard limit in label 1 we obtain:

\begin{equation}\label{hardeta}
    \eta_J \to \sum_{|\hat{I}|=k-1} h(e_{1\hat{I}}-e_J)s_{1\hat{I}}= \sum_{|\hat{I}|=k-1} h(e^{(1)}_{\hat{I}}-e_J)\hat{s}_{\hat{I}}\,.
\end{equation}

Recall that the hard limits $\hat{s}_{\hat{I}}$ are here interpreted as the kinematic invariants for $(k-1,n-1)$, i.e. a system that does not include particle label 1. From \eqref{hardeta} we conclude that the function $h(x-e_J)$ that defines de subdivision is restricted to the domain $x\in \partial^{(1)}\Delta_{k,n}$ after the hard limit is taken. Hence the boundary restriction of the subdivision is equivalent to the hard limit on the basis $\eta_J$, as we wanted to show.

Very nicely, the hard limit of $\eta_J$ also gives an element of the planar basis $\hat{\eta}_{\hat{I}}$ for $(k-1,n-1)$. For instance, if $J$ is of the form $J=(1\hat{J})$, it clear from \eqref{hardeta} that we obtain $\eta_{1\hat{J}} \to \hat{\eta}_{\hat{J}}$. A similar analysis can be done for general $J$ and recovers the general rule given in \cite{Early:2019eun} for the boundary map.

 A future direction is to further elucidate the relation between the compatibility formulation using Steinman relations/Weak Separation \cite{Early:2019zyi,Early:2019eun} and the compatibility criteria implemented here in the context of  planar arrays of degenerate Feynman diagrams.

\vskip0.1in

{\bf Note Added:}

While the second version of this paper was being prepared, some new results on local planarity appeared \cite{Cachazo:2022pnx,Cachazo:2023ltw}. In this paper, we have focused on the poles of the partial amplitude $ m^{(k)}_n  ( \mathbb{I}|\mathbb{I})$ consistent with a notion of global planarity. In \cite{Cachazo:2022pnx,Cachazo:2023ltw}, other kinds of partial amplitudes are introduced using the notion of local planarity, i.e.,  generalized color orderings, to define the color-dressed generalized biadjoint amplitudes.  It would be interesting to apply our method in this paper to study the more general partial amplitudes there
using their poles as degenerate arrays and their  compatibility criteria.


\section*{Acknowledgements}

We thank F. Cachazo and B. Gimenez for collaboration in the initial stages of this project. We would like to thank F. Cachazo and N. Early for useful discussions. Research at Perimeter Institute is supported in part by the Government of Canada through the Department of Innovation, Science and Economic Development Canada and by the Province of Ontario through the Ministry of Economic Development, Job Creation and Trade.


\bibliographystyle{JHEP}
\bibliography{references}

\end{document}